\documentclass[reprint, amsmath,amssymb, aps,prd,showpacs,
superscriptaddress,
]{revtex4-2}

\usepackage{graphicx}
\usepackage{dcolumn}
\usepackage{bm}
\usepackage{mathtools}
\usepackage[utf8]{inputenc} 
\usepackage{xcolor}
\usepackage{lineno}
\usepackage{lipsum}
\usepackage{aas_macros}
\usepackage{url}
\usepackage[colorlinks]{hyperref}
\usepackage{booktabs}
\usepackage{physics}

\AtBeginDocument{
\heavyrulewidth=.08em
\lightrulewidth=.05em
\cmidrulewidth=.03em
\belowrulesep=.65ex
\belowbottomsep=0pt
\aboverulesep=.4ex
\abovetopsep=0pt
\cmidrulesep=\doublerulesep
\cmidrulekern=.5em
\defaultaddspace=.5em
}

\definecolor{dodgerblue}{HTML}{1E90FF}
\definecolor{viennared}{HTML}{DA0A14}
\definecolor{ctorange}{HTML}{FF6C0C}
\definecolor{wales}{HTML}{ff0038}
\definecolor{benettongreen}{HTML}{009421}
\definecolor{ferrarired}{HTML}{ff2800}
\definecolor{austriawienpurple}{HTML}{441678}

\hypersetup{
     colorlinks=true,
     linkcolor=dodgerblue,
     filecolor=dodgerblue,
     citecolor = dodgerblue,      
     urlcolor=dodgerblue,
     }

\begin{document}
\preprint{APS/123-QED}

\title{Constraining the lensing of binary black holes from their stochastic background}

%
\author{Riccardo Buscicchio}
\email{riccardo@star.sr.bham.ac.uk}

\affiliation{School of Physics \& Astronomy, University of Birmingham, Birmingham, B15 2TT, UK
}
\affiliation{Institute for Gravitational Wave Astronomy, University of Birmingham, Birmingham, B15 2TT, UK
}
\author{Christopher J.\ Moore}

\affiliation{School of Physics \& Astronomy, University of Birmingham, Birmingham, B15 2TT, UK
}
\affiliation{Institute for Gravitational Wave Astronomy, University of Birmingham, Birmingham, B15 2TT, UK
}
\author{Geraint Pratten}
\affiliation{School of Physics \& Astronomy, University of Birmingham, Birmingham, B15 2TT, UK
}
\affiliation{Institute for Gravitational Wave Astronomy, University of Birmingham, Birmingham, B15 2TT, UK
}
\author{Patricia Schmidt}

\affiliation{School of Physics \& Astronomy, University of Birmingham, Birmingham, B15 2TT, UK
}
\affiliation{Institute for Gravitational Wave Astronomy, University of Birmingham, Birmingham, B15 2TT, UK
}
\author{Matteo Bianconi}
\affiliation{School of Physics \& Astronomy, University of Birmingham, Birmingham, B15 2TT, UK
}

\author{Alberto Vecchio}

\affiliation{School of Physics \& Astronomy, University of Birmingham, Birmingham, B15 2TT, UK
}
\affiliation{Institute for Gravitational Wave Astronomy, University of Birmingham, Birmingham, B15 2TT, UK
}

\date{\today}
\begin{abstract}
Gravitational waves (GWs) are subject to gravitational lensing in the same way as electromagnetic radiation. However, to date, no unequivocal observation of a lensed GW transient has been reported. 
Independently, GW observatories continue to search for the stochastic GW signal which is produced by many transient events at high redshift. We exploit a surprising connection between the lensing of individual transients and limits to the background radiation produced by the unresolved population of binary back hole mergers: we show that it constrains the fraction of individually resolvable lensed binary black holes to less than $\sim 4\times 10^{-5}$ at present sensitivity.
We clarify the interpretation of existing, low redshift GW observations (obtained assuming no lensing) in terms of their apparent lensed redshifts and masses and explore constraints from GW observatories at future sensitivity. Based on our results, recent claims of observations of lensed events are statistically disfavoured.
\end{abstract}

\maketitle
%
%
%
\noindent{\bf \em Introduction~--~}
Several binary black hole (BBH) mergers have been detected so far~\cite{2019PhRvX...9c1040A,LIGOScientific:2020stg} and a number of additional candidates reported ~\cite{2020ApJ...896L..44A, 2020ApJ...891..123N, 2019arXiv190407214V, 2019arXiv191009528Z}.
Forthcoming gravitational-wave (GW) detector upgrades will provide increased sensitivity, which will allow us to probe an even larger spacetime volume~\cite{2018LRR....21....3A}.

The current BBH detections are loud and individually resolvable~\cite{1993PhRvD..47.2198F, 1994PhRvD..49.2658C}.
However, they are part of a much larger population~\cite{2018LRR....21....3A} whose properties, such as the overall merger rate and the source mass distribution, can be inferred statistically~\cite{2019ApJ...882L..24A,2018ApJ...856..173T}. 
As new GW events are detected, this population can be constrained with increasing accuracy.
The GW ensemble redshift distribution and correlations with source parameters constitute an important piece of evidence, allowing us to place tighter constraints on progenitors formation history and evolution channels~\cite{2017ApJ...851L..25F, 2019MNRAS.484.4216R, 2017PhRvD..95l4046G, 2020arXiv200500023K, 2017PhRvD..96b3012T, 2018ApJ...863L..41F, 2018PhRvD..98h4036G, 2020arXiv200400650B}.
Ultimately, observing distinctive features in the population distribution would provide independent characterization of the expansion history of nearby universe~\cite{2019ApJ...883L..42F}.
Importantly, this population does not only consist of individually detectable BBH mergers but will contain many other distant, unresolved events~\cite{2019RPPh...82a6903C}.
Their emissions accumulate across all redshifts as a stochastic background of GWs (SGWB): an incoherent superposition of signals whose properties cannot be inferred individually~\cite{2017LRR....20....2R, 2020MNRAS.496.3281S}.

Broadly speaking, events are individually observable depending on the instrument sensitivities and the choice of search strategy \cite{2017PhRvD..95d2001M, 2016CQGra..33u5004U, 2016PhRvD..93d2004K, 2019MNRAS.484.4008G}.
The majority of events that are not individually observed contribute instead to the SGWB.
Current estimates predict a detection of a SGWB with a signal-to-noise ratio (SNR) of 3 after 40 months of observations \cite{2019PhRvD.100f1101A, 2018PhRvL.120i1101A}.
The observation of a stochastic background will complement individual detections, providing an integrated measure of the cosmological black holes' population history~\cite{2020arXiv200312152C}. 

GWs from BBHs are generated by the dynamics of vacuum spacetime, as prescribed by general relativity.
As a consequence, they carry information from an inherently scale-free physics.
Additional assumptions on the formation mechanism, or observations of a counterpart are necessary to connect with weak or electromagnetic phenomena thereby introducing new length/energy scales and breaking the ubiquitous mass-distance degeneracy~\cite{2019ApJ...883L..42F}.

However, GWs are in principle affected by the intervening gravitational potential which influences the inferred spatial and temporal properties of the signals~\cite{2010ARA&A..48...87T}. 
At the simplest level of description, the effect of lensing on a GW signal is to change its strain amplitude by a multiplicative magnification factor $\sqrt{\mu}$.
As a consequence, and in absence of independent constraints on the lensing magnification, the mass-distance degeneracy is re-established even for chirping sources.
Parameter estimation pipelines do not currently incorporate any lensing model, and therefore infer source properties agnostically of such a phenomenon.
However, follow-up studies have addressed a number of questions: Is any detection actually magnified? 
Are there event couples originating from the same source emission, whose light-path has been altered to mimic independent events? 
Does lensing affect the population inference?~\cite{2018IAUS..338...98S, 2019ApJ...874L...2H, 2019MNRAS.485.5180S, 2018MNRAS.475.3823S, 2017PhRvD..95d4011D, 2018PhRvD..97b3012N, 2018MNRAS.476.2220L, 2019arXiv190406020L, 2018arXiv180205273B, 2019arXiv190103190B, 2020arXiv200208821B}. 

In this letter we address one of the above questions, rephrasing it as a probabilistic statement. 
Given a set of observations, how likely is it for a fraction to be magnified by more than a certain $\mu$? 
We show that by considering lensing of the entire population a significant amount of information can be leveraged from the SGWB; even the current non-detection has surprising astrophysical consequences. Significantly, we find the recent claims of lensed events to be statistically disfavoured~\cite{2018arXiv180205273B, 2019arXiv190103190B, 2020arXiv200208821B}.

%
%
\noindent{\bf \em Modeling~--~}
We now turn our attention to the modeling assumptions made.
Firstly, we describe the effect of lensing on GW signals, and the parameterization of the lensing probability model.
Secondly, we summarize the features of the population model for BBH mergers. 
Finally, we derive the associated energy density of the stochastic background including lensed events.
Throughout this letter we use $G=c=1$.

%
%
\noindent{\em Lensing-probability~--~}
Unlensed, chirping binaries provide a direct measurement of their luminosity distance $d_L$~\cite{1986Natur.323..310S,2005ApJ...629...15H}. 
If associated with electromagnetic counterparts, this gives an independent estimate of the source redshift $z$.
Together, these constitute a point measurement in the expansion history of the universe~\cite{2018Natur.562..545C, 2019arXiv190806060T}
\begin{align}
    \frac{d_L(z)}{1+z}= \frac{1}{H_0} \int_0^z\text{d}z^\prime \frac{1}{E(z^\prime)}\,,
\end{align}
where $H_0$ is the local Hubble constant. $E(z)$ is a function of redshift, proportional to the time derivative of the logarithm of the scale factor, and encodes the information on the cosmological density parameters.

Alternatively, assuming a cosmological model breaks the mass-redshift degeneracy, thereby providing a redshift estimate for each observed event. 
However, this degeneracy is re-established by the addition of an \emph{a priori} unknown lensing magnification $\mu$.

Given a GW event, we focus on its true luminosity distance $d_L(z)$, chirp mass $\mathcal{M}$, and lensing magnification $\mu$. 
Its strain amplitude is magnified by a multiplicative factor $\sqrt\mu$~\cite{2010ARA&A..48...87T}.
Independent of the cosmology, the apparent mass $\tilde{\mathcal{M}}$, redshift $\tilde{z}$, and distance $\tilde{d}_L$ are related to their true values by the following relationships:
\begin{align}\label{eq:degeneracy}
    \frac{d_L(\tilde{z})}{\sqrt{\tilde{\mu}}}=\frac{d_L(z)}{\sqrt{\mu}}\;\;,\;\; \quad
    \tilde{\mathcal{M}}(1+\tilde{z})=\mathcal{M}(1+z)\;.
\end{align}
The apparent parameters are those inferred by any pipeline that \emph{assumes} a certain magnification $\tilde{\mu}$.
Parameter estimates provided in published catalogues are computed under the assumption of no lensing, i.e. $\tilde{\mu}=1$~\cite{2019PhRvX...9c1040A,2020ApJ...891..123N, 2019arXiv190407214V, 2019arXiv191009528Z}.

In order to incorporate the effect of lensing in the parameter reconstruction, additional independent information on the same transient would be required: e.g. the observation of electromagnetic counterparts, a detailed knowledge of the lensing potential along the GW travel path, or an association with a host galaxy.
Another possibility is the association between two or more GW events, whose apparent properties can be referred back to a common source that has undergone multiple imaging~\cite{2019ApJ...874L...2H}. 
In the absence of such additional information, prior knowledge on $\mu$ remains unaltered after any single detection, because of the above degeneracy.

In this letter we use a semi-analytic \emph{lensing model} for the probability of a given magnification $\mathrm{d}P/\mathrm{d}\ln\!\mu$ from equation ($\mathrm{B1}$) in~\cite{2017PhRvD..95d4011D}.
We use cubic splines to interpolate the data from Table $\mathrm{I}$ in~\cite{2017PhRvD..95d4011D} across the redshift range $z \in \left[0,20\right]$.
We note that this model correctly captures the limiting behavior in both the strong and weak lensing regimes~\cite{2008MNRAS.386.1845H,2011ApJ...742...15T}, and is in agreement with recent hydrodynamical simulations~\cite{2020MNRAS.tmp.1583R}.
The lensing model is described in more detail in the Supplemental Material.

%
%
\noindent{\em BBH populations~--~}
Following ~\cite{2018ApJ...863L..41F, 2019ApJ...882L..24A} (which are based on~\cite{2019PhRvX...9c1040A}), we parametrize the BBH differential merger rate $R$ as a function of the binary masses $m_{1,2}$ and redshift $z$, as
\begin{align}\label{eq:rate-density}
    \frac{\text{d}^3R}{\text{d}m_1 \text{d}m_2 \text{d}z}\!=\! \mathcal{R}(z|\lambda,\!\gamma,\!z_\mathrm{P}) p(m_1,\!m_2|m_\text{min},m_\text{max},\alpha),
\end{align}
where
\begin{align}\label{eq:mass-distro}
    p(m_1,m_2\mid m_\text{min},m_\text{max},\alpha) &\propto m_1^{-\alpha}& \\ 
    &\times \mathbf{I}(m_1\mid m_\text{min},m_\text{max}) \nonumber\\
    &\times \mathbf{I}(m_2\mid m_\text{min},m_1) \,.\nonumber
\end{align}
Here $\mathbf{I}(\cdot|a,b)$ are the indicator functions on the interval $[a, b]$, and throughout we adopt $\alpha\!=\!2.3$ and $(m_{\rm min},m_{\rm max})\!=\!(5,50)M_\odot$.

We set the cosmological merger rate to track the star formation rate (with no delay between formation and coalescence). 
This is modeled using a power law with index $\lambda$ peaking at $z_\mathrm{P}$ and tapering off further in the past with index $(\lambda-\gamma)$ \cite{2014ARA&A..52..415M};
\begin{align}\label{eq:rate-redshift}
    \mathcal{R}(z\mid\lambda, \gamma, z_\mathrm{P}) = R_0 \dfrac{(1+z)^\lambda}{1+\left[\dfrac{1+z}{1+z_\mathrm{P}}\right]^\gamma} \,.
\end{align}

\begin{table}[t]
\centering
\begin{tabular}{l | c c c c }
\toprule
 & $R_{0} \, (\text{Gpc}^{-3}\text{yr}^{-1})$ & $\lambda$ & $\gamma$ & $z_\mathrm{P}$\\ \midrule
$\mathrm{O1+O2}$ & $57^{+40}_{-25}$ & $5.8^{+0.4}_{-0.4}$ & $5.6$ & $1.9$ \\
Design & $57^{+40}_{-25}$ & $3.4^{+0.6}_{-0.7}$ & $5.6$ & $1.9$ \\ \bottomrule
\end{tabular}
\caption{\label{tab:models}Parameters modeling the merger rate density. $R_0$ is tuned to match the current estimate of the local merger rate, while $z_\mathrm{P},\gamma$ capture the the star formation rate peak and decay further out in redshift. $\lambda$ is adjusted to provide a stochastic background signal with a fixed SNR=2 at the two sensitivities considered. Fig.~\ref{fig:madau-dickinson} shows the two resulting distributions.
}
\end{table}

\begin{figure}[t]
\includegraphics[width=1\columnwidth]{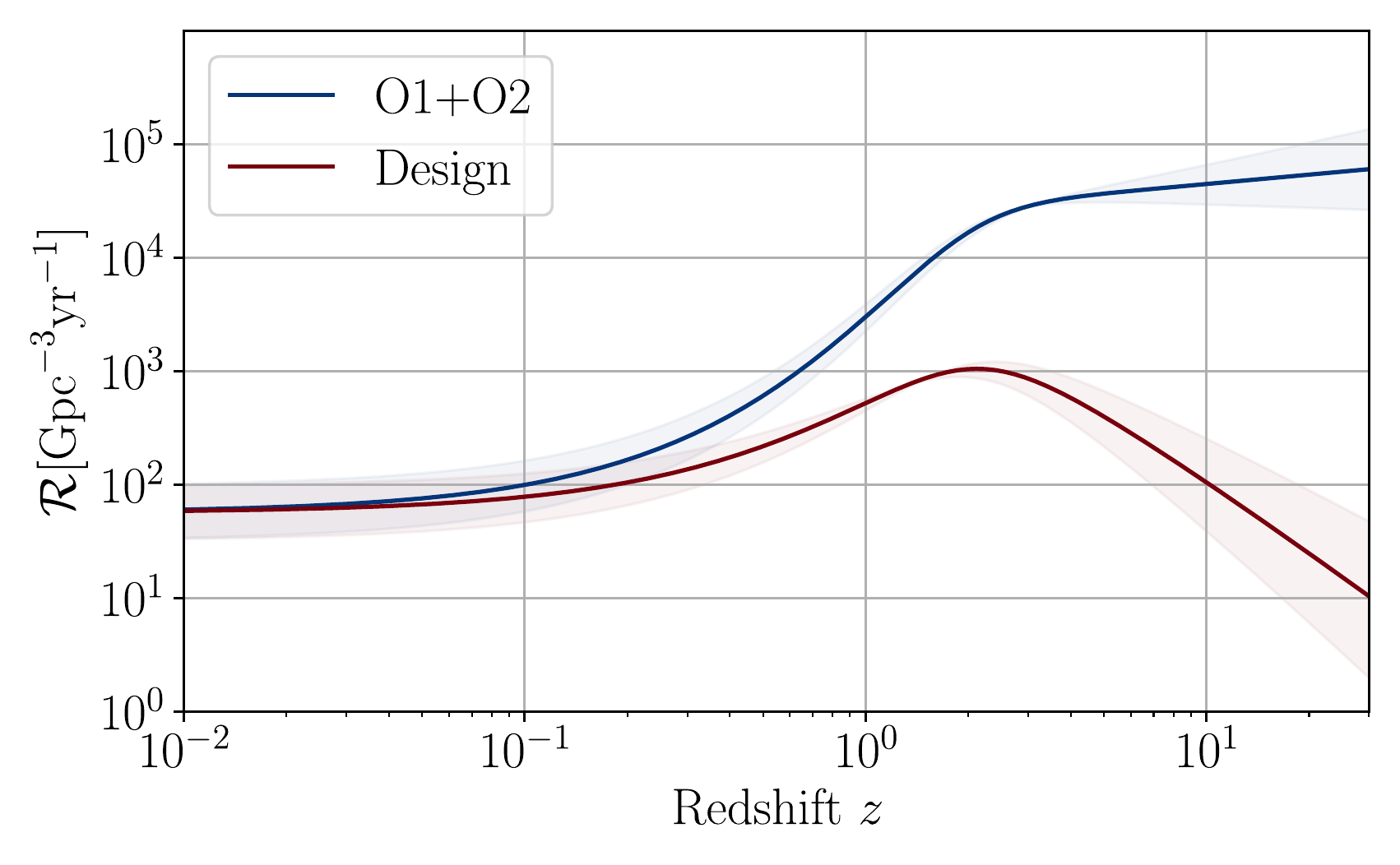}
\caption{\label{fig:madau-dickinson}
Cosmological merger rate density models considered, using the parameterization in Eq.~\ref{eq:rate-redshift}. 
Parameter choices are listed in Table~\ref{tab:models}. 
Models are matched to the current local estimate for the BBH merger rate, and to a stochastic signal with SNR=2 at a given sensitivity. 
The blue line refers to the sensitivity achieved after O1 and O2.
The red line refers to the projected sensitivity after two years of observation at $50\%$ duty cycle at design sensitivity. 
Shaded regions delimit analogous models, tuned to the upper and lower $90\%$ confidence interval on the local merger rate estimate.}
\end{figure}

We tune $R_0$ to match the current estimate for the local merger rate from GW population analyses~\cite{2019ApJ...882L..24A}.
The uncertainty on the merger rate propagates to all redshifts affecting the entire population.

A straightforward consequence of fixing the mass and redshift distributions (Eqs.~\ref{eq:mass-distro} and \ref{eq:rate-redshift} respectively) is the amplitude of the stochastic background accumulated over the past light-cone of the observer.
We have considered a number of merger rate models, varying both $\lambda$ and $\gamma$, while keeping the star formation rate peak $z_\mathrm{P}$ fixed.
In this letter, we present results for two choices of $\lambda$, with a fixed $\gamma$, that yield SGWB amplitudes consistent with current upper limits~\cite{2020arXiv200312152C} (see Fig.~\ref{fig:madau-dickinson}, and Table~\ref{tab:models}, and the discussion in the following section).

For simplicity, we neglect in both models black hole spins.
Depending on the spin properties of the BBHs, the enhancement on the overall rate can be significant, up to a factor of 3 in the mass range of interest for current detectors~\cite{2018PhRvD..98h4036G}.
We leave a consistent inclusion of spin effects --i.e. on the intrinsic merger rate, on the spectral shape of the stochastic signal, and on individual event detectability-- to future work.

We highlight that the local merger rate and mass distribution used here were obtained with hierarchical analyses on individual GW source parameters~\cite{2010ApJ...725.2166H,2010PhRvD..81h4029M,2019MNRAS.486.1086M}.
Therefore, in order to remain consistent with the prior assumptions therein, we have to consider $z, m_1, m_2$ as the {\em apparent} redshifts and masses with no intervening lensing, i.e. assuming $\tilde{\mu}=1$.
Henceforth, building on the notation in Eq.~\ref{eq:degeneracy}, we denote these parameters $\tilde{z}, \tilde{m}_1, \tilde{m}_2$, and related functions with a superscript tilde.
%
%

\noindent{\em Lensed stochastic background~--~}
The stochastic background from BBH mergers is the incoherent superposition of individual GW events~\cite{2001astro.ph..8028P, 2018PhRvL.120i1101A}.
We assume a flat $\Lambda\text{CDM}$ cosmology and a simple leading order post-Newtonian expression for the GW energy spectrum from the inspiral of non-spinning BBHs~\cite{1963PhRv..131..435P}
\begin{align}
    \frac{\text{d}E_\text{GW}(m_1,m_2)}{\text{d}f_r}\,=\,\frac{(\pi)^{2/3}}{3}\mathcal{M}(m_1,m_2)^{5/3}f_r^{-1/3}\,,
\end{align}
with $f_r=f(1+z)$ the GW frequency in the source rest frame.
Integrating over the cosmological expansion history, gives a value for the energy density of GWs from BBH mergers expressed as a fraction of the critical density $\rho_c$: 
this is a standard result from the GW literature\cite{2001astro.ph..8028P}, and it reads

\begin{align}\label{eq:unlensed-omega}
\Omega_\text{BBH}(f)&=\frac{1}{\rho_c}\int\text{d}z \frac{f}{H_0(1+z)E(z)}\\ &\times\int\text{d}m_1\text{d}m_2\frac{\text{d}^3R}{\text{d}m_1 \text{d}m_2 \text{d}z}\frac{\text{d}E_\text{GW}}{\text{d}f_r}\Bigg|_{f_r=f(1+z)}\,.\nonumber
\end{align}

We stress here that Eq.~\ref{eq:unlensed-omega} neglects the effects of lensing. 
Here, we seek to instead compute $\tilde{\Omega}_\mathrm{BBH}$ which accounts for the lensing model.
In order to do this we must modify Eq.~\ref{eq:unlensed-omega} by replacing  $\{\mathrm{d}z,\mathrm{d}m_{1,2}\}\rightarrow\{\mathrm{d}\tilde{z},\mathrm{d}\tilde{m}_{1,2}\}$, use the apparent differential merger rate $\mathrm{d}^3\tilde{R}$, and use the apparent redshifted frequency $f_r=f(1+\tilde{z})$.

We constrain the maximum allowed redshift evolution -- $\lambda$ in Eq.~(\ref{eq:rate-redshift}) -- by considering upper-limits on a SGWB~\cite{2019PhRvD.100f1101A,2013PhRvD..88l4032T}, while keeping the local merger rate fixed to the observed value.
We consider the current SGWB limit based on the O1 and O2 observing runs, using data from the two LIGO instruments only. As a limit for a non-detection we assume a signal-to-noise ratio smaller than 2 in a stochastic search~\cite{2013PhRvD..88l4032T}.
Similarly, we forecast the projected limits after $2$ years of observation at design sensitivity and $50\%$ duty cycle of the network of the two LIGO instruments and Virgo. We denote the two scenarios \emph{O1+O2} and \emph{Design}, respectively.

As expected and clearly shown in Fig.~\ref{fig:madau-dickinson} a non detection of a SGWB over longer integration time and with better sensitivities implies a lower merger rate outside the horizon for individual detections. The merger rate redshift evolution considered here is consistent with the results of~\cite{2020arXiv200312152C}.

%
%
\noindent{\bf \em Lensing-fraction~--~}
Having established the population models to be considered, we turn to our main task: quantifying the probability for an individual transient to be magnified with a particular magnification.

\begin{figure}[t]
\includegraphics[width=1\columnwidth]{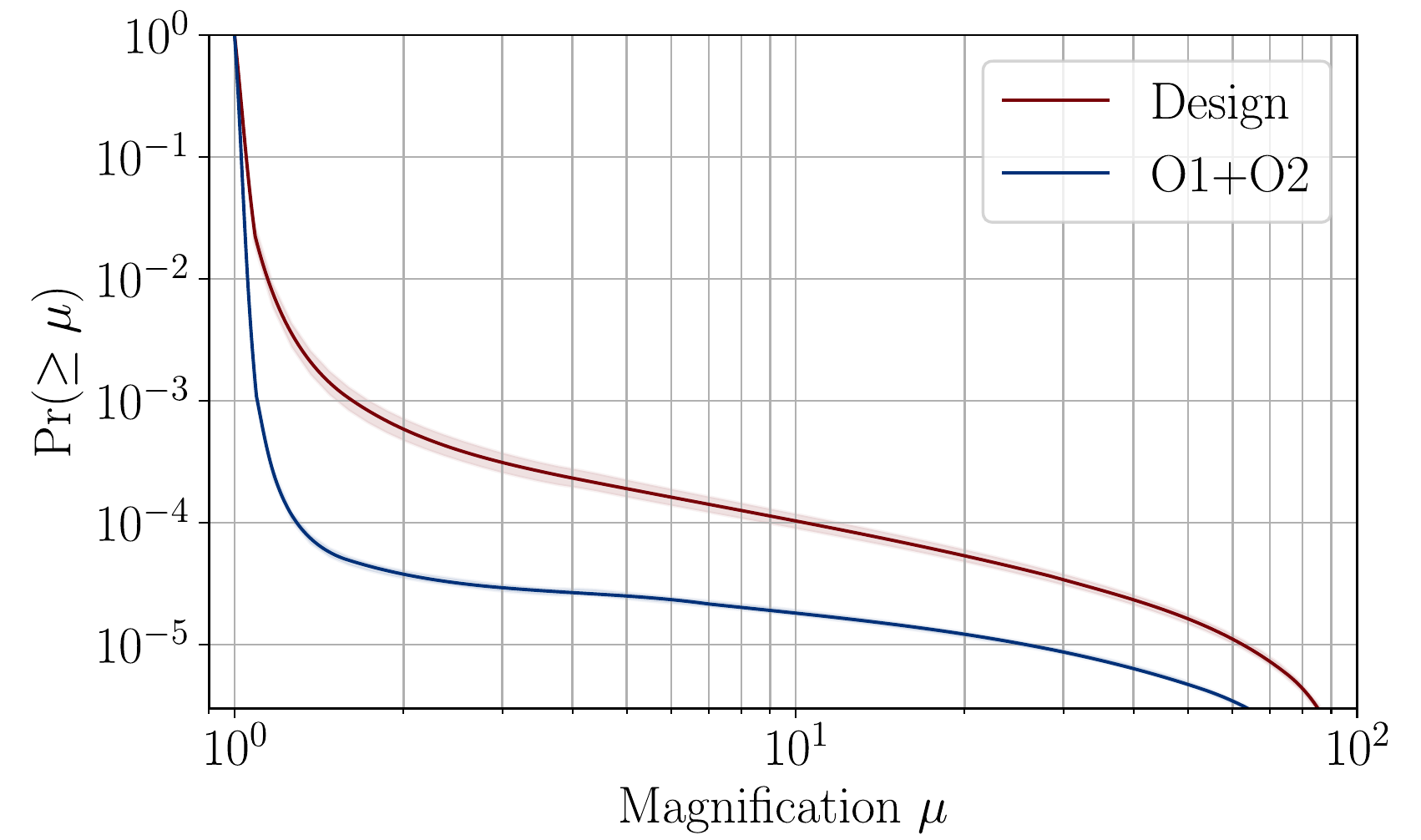}
\caption{\label{fig:lensing_fraction}
Complementary cumulative distribution for the lensing probability of detectable BBH mergers, constrained by the non-detection of the SGWB for two sensitivity scenarios. Solid lines and narrow shaded regions are obtained from corresponding models shown in Fig.~\ref{fig:madau-dickinson}. The fraction of lensed transients with $\mu>2$ is less than ${\sim 4\times10^{-5}}$ after O1 and O2; a non detection of a SGWB after 2 years of operation at design sensitivity would yield a fraction a factor of 10 higher. The result depends very weakly on the local merger rate uncertainty, hence the light-blue shaded region has neglible width.}
\end{figure}
\begin{figure*}[t]
\includegraphics[width=\textwidth]{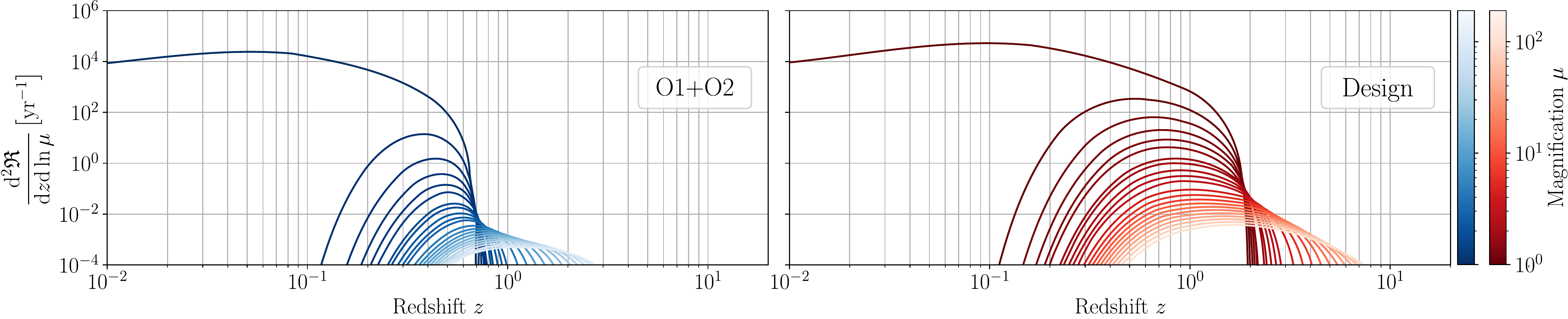}
\caption{\label{fig:integrands}
Differential rate of detectable lensed events for each redshift and logarithmic magnification bin.  
Results are shown as solid lines coloured according to magnification. 
Left (right) panels show results for the O1 and O2 (Design) population models respectively; see Table~\ref{tab:models} and the accompanying discussion in the text.
Moderately magnified events (e.g. $\mu<10$) dominates the detected population of BBH mergers by at least three orders of magnitude. At Design sensitivity, a non-detection of a stochastic background will imply by itself a significant reduction of mergers at high redshift, as described by the model in Fig.~\ref{fig:madau-dickinson}.
Concurrently, a better sensitivity enhances detections further out in redshift, at all magnifications. Predominantly, non magnified events will be observed out to $z\approx2$. A very small fraction of strongly magnified ones will extend out $z\approx6$.  
A comparison of the overall improvement integrated over redshift at each magnification, is presented in Fig.~\ref{fig:lensing_fraction}.
}
\end{figure*}
For each apparent redshift shell $\left[\tilde{z},\tilde{z}+\mathrm{d}\tilde{z}\right]$
we consider contributions from true redshifts shells up to $z=20$, the maximum extent of the lensing model~\cite{2017PhRvD..95d4011D}.
The pair $(z,\tilde{z})$ fixes uniquely the magnification and therefore the relationship between the two redshifts is given by Eq.~\ref{eq:degeneracy} (with $\tilde{\mu}=1$); this transformation, and its Jacobian, $|\partial \tilde{z}/\partial z|_{\mu}$, is further explored in the supplemental material.

We use this relationship to write explicitly an expression for the differential rate of magnified events which is a proxy for the magnification probability, 
\begin{align}\label{eq:lensed-fraction}
\frac{\mathrm{d}^{2}\mathfrak{R}}{\mathrm{d}z\mathrm{d}\ln\mu}&=\frac{\mathrm{d}P(\mu\mid z)}{\mathrm{d}\mathrm{\ln\mu}} \frac{4\pi \chi^{2}\left(z\right)}{H_{0}\left(1+z\right)E\left(z\right)}\left|\frac{\partial\tilde{z}}{\partial z}\right|_{\mu}\\ 
&\times\int \mathrm{d}\tilde{m}_{1}\mathrm{d}\tilde{m}_{2}\frac{\mathrm{d}^{3}\tilde{R}}{\mathrm{d}\tilde{m}_{1}\mathrm{d}\tilde{m}_{2}\mathrm{d}\tilde{z}}p_{\mathrm{det}}\left(\tilde{m}_{1},\tilde{m}_{2},\tilde{z}\right)\nonumber \,.
\end{align}
Additionally, using apparent masses and redshifts we filter events by their detectability. We use a fixed single detector threshold $\mathrm{SNR}=8$ for each given set of source parameters, and compute the observable fraction of the distribution in component masses, averaged over the source orientation~\cite{1993PhRvD..47.2198F}. We estimate selection effects $p_\mathrm{det}(\tilde{m}_1,\tilde{m}_2,\tilde{z})$ for both sensitivities using the publicly available code \textsc{gwdet}~\cite{gwdet}. 

As discussed above, the SGWB should contain only unresolved events. To be consistent, the same selection effects should be added to Eq.~\ref{eq:unlensed-omega} by the inclusion of a factor $(1\!-\!p_\mathrm{det}(\tilde{m}_1,\tilde{m}_2,\tilde{z}))$ in the innermost integral.
However, we neglect this effect here because the region of parameter space where $p_\mathrm{det}$ is non-zero, i.e.\ at moderate masses and low redshift, is far from the peak of the intrinsic rate.

Results are shown in Figs.~\ref{fig:lensing_fraction} and \ref{fig:integrands}. 
Detections are dominated in both scenarios by low-redshift, unlensed events (i.e. $\mu\approx 1$). 
While at design sensitivity the detections will extend further out to $z\approx2$, magnifications smaller than 2 will likely dominate the population by at least three orders of magnitude.
This is clearly apparent in Fig.~\ref{fig:lensing_fraction}, where the contributions across redshifts are integrated out to a single magnification distribution. 
For ease of comparison we show both as cumulative distribution functions, i.e. factoring out the respective total rate of detections per year.

Remarkably, a better instrument sensitivity provides proportionally more events at larger magnification. 
This is the net result of a few competing factors.
The assumed non-detection of a SGWB constrains the population to a shallower redshift distribution: 
as a consequence both lensed and unlensed events within the detection horizon are equally suppressed;
however, the population of distant events at $z>2$ is significantly depleted, therefore reducing their relative contribution to the apparent distribution.

We study the impact on our results of our modelling choices for (i) the lensing model, (ii) the redshift evolution of the merger rate, (iii) the BBH mass distribution.
Overall we find our results to be robust; changing the mass distribution has the largest effect, increasing the fraction of lensed event by at most a factor of 2 (see details in Supplemental material).

%
%
\noindent{\bf \em Conclusions~--~}
A SGWB of astrophysical origin has not yet been observed.
This constrains the redshift dependence of the BBH merger rate, particularly the number of mergers at high redshift. 
This in turn has consequences for the lensing probability of individual events.

In this letter we exploit this surprising link between the non detection of a SGWB and the lensing probability to quantify the fraction of lensed BBH events.
We provide estimates for the relative contribution of lensed BBHs to the total rate out to redshifts of $z\leq 20$ and magnifications of $\mu\leq 100$.
Even the current non-detection of a SGWB already has interesting astrophysical implications; we find a fraction below $\sim 4\times 10^{-5}$
of events to have a magnification $\mu\geq 2$.
At design sensitivity, in the absence of a SGWB detection after two years of observation, this fraction increases by a factor of $\sim 10$.

If and when there is a detection of a SGWB, our argument will become even more informative.
It can be applied to the BBH merger redshift distribution --constrained jointly from the mergers population and the SGWB detection-- to predict the number of lensed events. 
For a detection of a SGWB in less than two years of observation at \emph{Design} sensitivity, we expect the inferred lensing fraction to lie between the two curves shown in Fig.~\ref{fig:lensing_fraction}.

Simultaneously and independently, a similar study using complementary methods appeared~\cite{2020arXiv200603064M} showing agreement with our results.

%
%
\noindent{\bf \em Acknowledgments~--~}
The authors thank Davide Gerosa and Graham P. Smith for useful comments. RB thanks Will M. Farr, Thomas Callister and Katerina Chatziioannou for fruitful discussions.
AV acknowledges the support of the Royal Society and Wolfson Foundation. PS acknowledges NWO Veni Grant No. 680-47-460.
Computational work was performed using the University of Birmingham's BlueBEAR HPC service.
\vfill
\bibliographystyle{apsrev4-2}
\bibliography{biblio}
\vfill

\end{document}


\preprint{APS/123-QED}

\clearpage
\widetext
\begin{center}
  \textbf{\large Supplemental Material:} \\
\end{center}
\begin{center}
    \textbf{Relating the true and apparent redshifts}
\end{center}
\setcounter{equation}{0}
\setcounter{figure}{0}
\setcounter{table}{0}
\setcounter{page}{1}
\makeatletter
\renewcommand{\theequation}{S\arabic{equation}}
\renewcommand{\thefigure}{S\arabic{figure}}
\renewcommand{\bibnumfmt}[1]{[S#1]}
\renewcommand{\citenumfont}[1]{S#1}

\begin{figure}[hb]
\includegraphics[width=1\columnwidth]{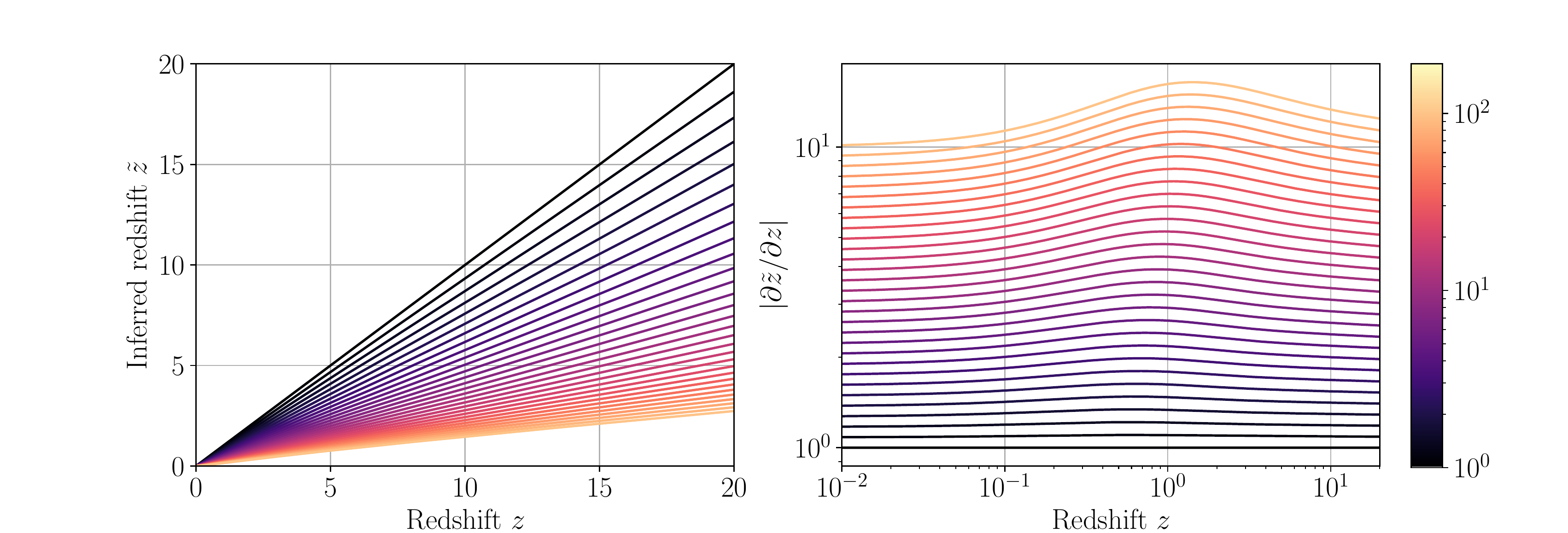}
\caption{\label{fig:apparent-vs-real}
Left panel: the apparent, or inferred redshift $\tilde{z}$ as a function of the true redshift $z$ for different magnifications; see \ref{eq:ztilde_supp}. 
Right panel: the Jacobian $|\partial\tilde{z}/\partial z|_{\mu}$ of the transformation; see \ref{eq:Jacobian_supp}.}
\end{figure}

As stated in the main text, when computing the fraction of lensed event, for each apparent redshift shell $\left[\tilde{z},\tilde{z}+\mathrm{d}\tilde{z}\right]$
we consider contributions from true redshifts shells up to the maximum extent of the lensing model~\cite{2017PhRvD..95d4011D}, i.e. $z\leq 20$.
The true and apparent redshift couple $(z,\tilde{z})$ fixes uniquely the magnification and therefore relationship between the two redshifts.
\begin{eqnarray}
    \tilde{z}&=&d_L^{-1}\left(\frac{d_L(z)}{\sqrt{\mu}}\right) \label{eq:ztilde_supp}\\
    \Rightarrow\left|\frac{\partial \tilde{z}}{\partial z}\right|_\mu&=&\sqrt{\mu}\frac{d_L^\prime(z)}{d_L^\prime(d_L^{-1}(\sqrt{\mu}d_L(z)))}\label{eq:Jacobian_supp}
\end{eqnarray}
The transformation \ref{eq:ztilde_supp} comes from rearranging Eq.1 from the main text, and in \ref{eq:Jacobian_supp} we give its Jacobian.
Both relationships are plotted in the two panels of Fig.~\ref{fig:apparent-vs-real}.
These quantities are used to write explicitly an expression for the rate of events per unit time (Eq.~8 of the main text) of individual GW transients, per unit redshift, per unit log-magnification;
\begin{align}\label{eq:lensed-fraction_supp}
\frac{\mathrm{d}^{2}\mathfrak{R}}{\mathrm{d}z\mathrm{d}\ln\mu}&=\frac{\mathrm{d}P(\mu\mid z)}{\mathrm{\ln\mu}} \frac{4\pi \chi^{2}\left(z\right)}{H_{0}\left(1+z\right)E\left(z\right)}\left|\frac{\partial\tilde{z}}{\partial z}\right|_{\mu}\int d\tilde{m}_{1}d\tilde{m}_{2}\frac{\mathrm{d}^{3}\tilde{R}}{\mathrm{d}\tilde{m}_{1}\mathrm{d}\tilde{m}_{2}\mathrm{d}\tilde{z}}p_{\mathrm{det}}\left(\tilde{m}_{1},\tilde{m}_{2},\tilde{z}\right)\,.
\end{align}
In Eq.~\ref{eq:lensed-fraction_supp} the lensing probability from Eq.~$\mathrm{B1}$ in~\cite{2017PhRvD..95d4011D} is multiplied by the spherical redshift shell -- parameterized here by its comoving radius $\chi(z)$--, the redshift transformation Jacobian in Eq.~\ref{eq:Jacobian_supp}, and the merger rate density in Eq.~3, integrated over the apparent mass distribution Eq.~4.
\begin{center}
  \textbf{Lensing model} \\
\end{center}

\begin{figure}[htb]
\includegraphics[trim=-2cm 0 0 0,width=.625\columnwidth]{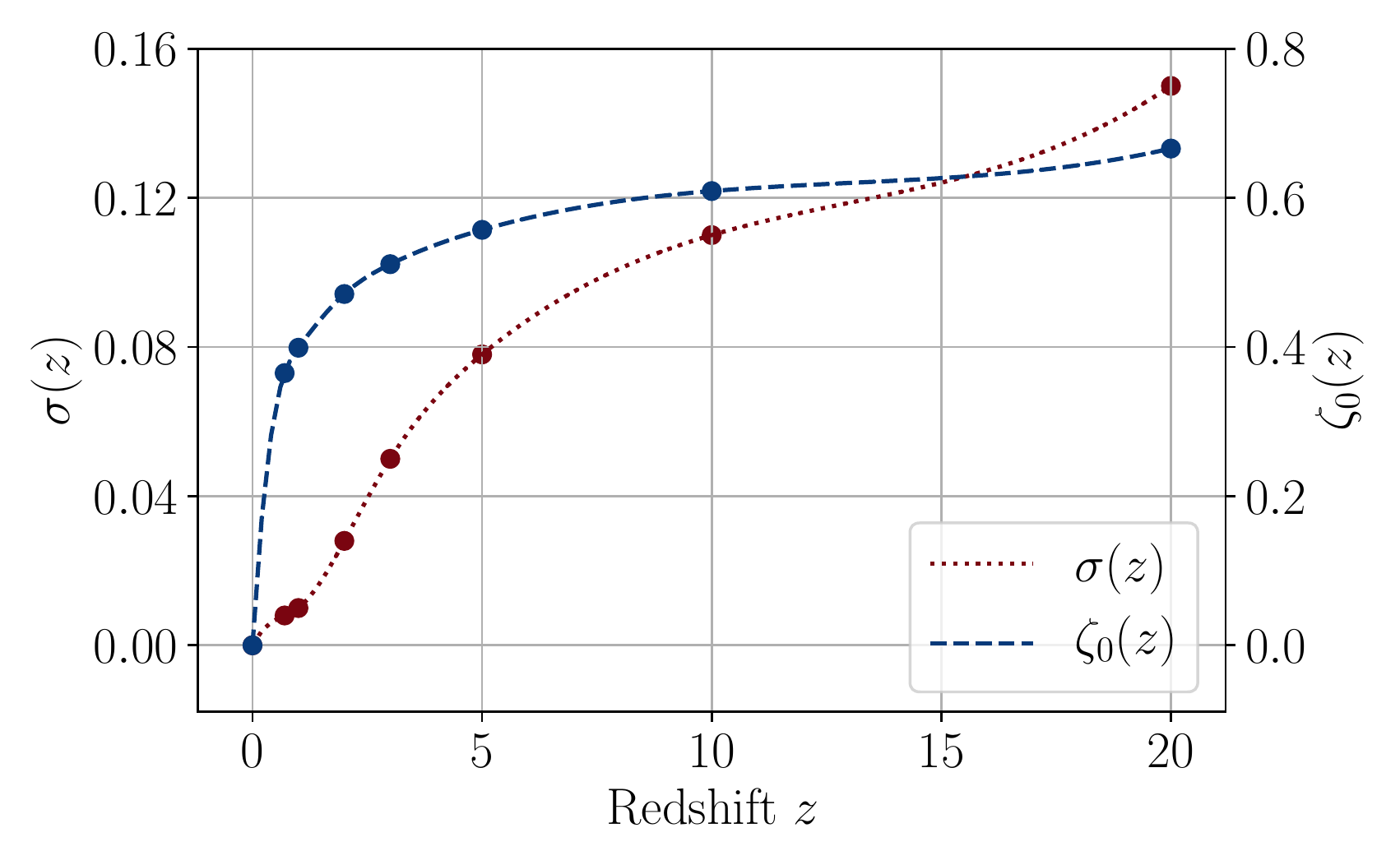}
\caption{\label{fig:interpolate-lensing-fits}
Hyperparameters for the lensing probability in Eq.~\ref{eq:lensing-probability}. Red (blue) dots describes best-fitting parameters for the log-normal kernel (exponential tail) of the magnification distribution. Values are interpolated across redshifts and match both strong and weak lensing predictions from ray-tracing simulations~\cite{2011ApJ...742...15T,2008MNRAS.386.1845H}.}
\end{figure}

We consider a semi-analytic parametric model for the lensing magnification~\cite{2017PhRvD..95d4011D} due to intervening galaxies.
The conditional differential probability for $\ln{\mu}$ reads
\begin{align}\label{eq:lensing-probability}
    \frac{\text{d}P(\ln\mu\!\mid\! z)}{\text{d}\ln\mu}\!\propto\!\int_0^{\infty}\!\text{d}\zeta\exp\!\left[\frac{5}{\zeta+\zeta_0}\!-2\zeta\right]\!
    \exp\left[-\left(\frac{\ln\mu-\delta-\zeta}{\sigma}\right)^2\right]
    \,,
\end{align}
with the three parameters $(\delta, \sigma, \zeta_0)$ functions of $z$. 
Assuming a mean magnification of 1 uniquely fixes the shift parameter, $\delta$. 
The remaining parameters, $\sigma$ and $\zeta_0$, are chosen as a convenient parameterization to fit the low and high magnification regimes of the distribution.
As shown in Fig.~\ref{fig:interpolate-lensing-fits}, we interpolate across redshift the available parameter values using cubic splines to obtain a smooth model of the magnification probability for redshifts $z \in \left[0,20\right]$.

As discussed in~\cite{2017PhRvD..95d4011D}, for large $\mu$ the gaussian asymptotes to a delta function, therefore the integral becomes proportional to $\mu^{-2}$. 
The normalization --dependent on $\zeta_0$-- is matched to predictions from ray-tracing studies on the Millenium Simulation~\cite{2008MNRAS.386.1845H}.
More recently, investigations through hydrodynamical simulations have shown consistent results on the high magnification lensing from galaxies and clusters dark matter halos~\cite{2020MNRAS.tmp.1583R}.
At the same time, the normal distribution in $\ln\mu$ is fixed through $\sigma$ to weak lensing predictions~\cite{2011ApJ...742...15T}.

\begin{center}
  \textbf{Impact of modelling assumptions} \\
\end{center}
To assess the stability of the results in the main text with respect to the modelling choice made, we allow for a number of variations: 
the peak $z_p$ in the interval $[1.0, 2.5]$ (compatible with the prior choice in~\cite{2020arXiv200312152C}); 
the lensing model to vary within the uncertainty of the semi-analitic fit in~\cite{2017PhRvD..95d4011D}; 
the mass distribution in Eq.4 of the main text is replaced by a broader \emph{flat-in-log} one, with binary component masses independently distributed as $\propto 1/m_i$ within the same range $[5\mathrm{M_\odot}, 50\mathrm{M_\odot}]$.
\begin{center}
    \begin{figure}[bh]
    \includegraphics[width=0.5\columnwidth]{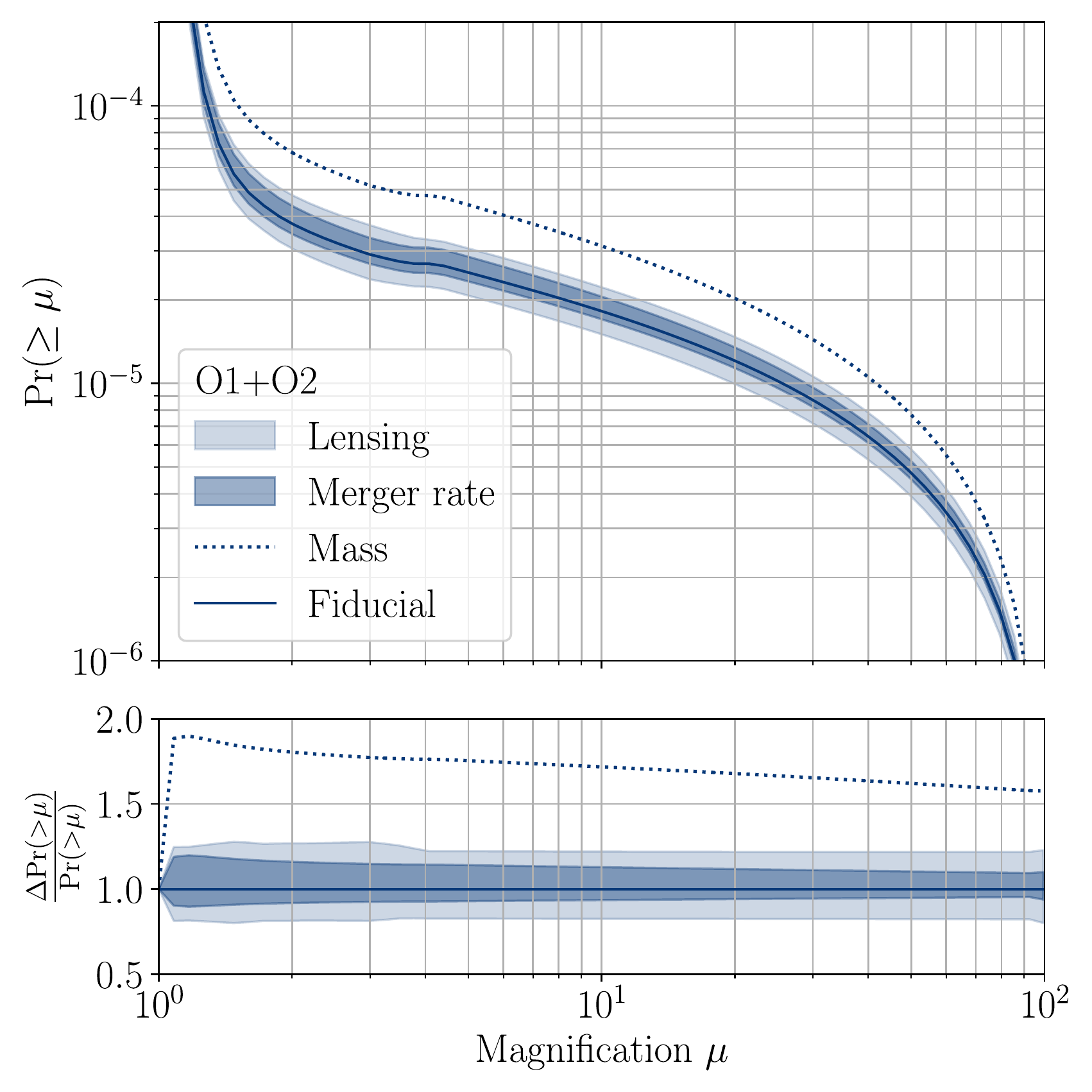}
    \caption{\label{fig:stability} Impact of different modeling assumptions onto the resulting lensing fraction after O1 and O2.
    Top panel: Solid (dotted) lines correspond to the powerlaw (flat-in-log) mass distribution presented in the main text (this Supplement). Light shaded regions correspond to variation merger rate density peak $z_p$. Dark shaded regions correspond to variation in the lensing model within the uncertainties of the numerical fit.
    Bottom panel: Ratio of every model presented in the top panel with respect to the one presented in the main text (labelled here \emph{Fiducial})
    Overall the variation in the resulting lensing fraction are smaller than a factor of two for both sensitivities.}
    \end{figure}
\end{center}
As shown in Fig.~\ref{fig:stability} we observe a variation in the resulting lensing fraction of less than a factor of $2$ at each magnification at \emph{O1+O2} sensitivity. A similar effect is observed at \emph{Design} sensitivity.
\bibliographystyle{apsrev4-2}
\bibliography{biblio}

\vfill